\documentclass[aps,reprint]{revtex4-2}

\usepackage{amsmath}
\usepackage{graphicx}
\usepackage{chemformula}
\usepackage{subcaption}
\usepackage{siunitx}

\graphicspath{{figures/}}

\renewcommand{\vec}[1]{\mathbf{#1}}
\newcommand{\abs}[1]{\left\vert #1 \right\vert}

\begin{document}

\title{Dexterous holographic trapping of dark-seeking particles with
Zernike holograms}

\author{Jatin Abacousnac}
\author{David G. Grier}

\affiliation{Department of Physics 
and Center for Soft Matter Research, 
New York University, 
New York, NY 10003, USA}

\begin{abstract}
The intensity distribution of a
holographically-projected optical trap
can be tailored to
the physical properties of the particles
it is intended to trap.
Dynamic optimization is especially desirable
for manipulating dark-seeking
particles that are repelled by conventional optical
tweezers, and even more so when 
dark-seeking particles coexist in
the same system as light-seeking particles.
We address the need for dexterous manipulation
of dark-seeking particles by introducing
a class of ``dark'' traps created from
the superposition of two out-of-phase Gaussian modes
with different waist diameters.
Interference in the difference-of-Gaussians (DoG) trap
creates a dark central core that is completely surrounded
by light and therefore can trap
dark-seeking particles rigidly in three dimensions.
DoG traps can be combined with conventional optical
tweezers and other types of traps for use in heterogeneous
samples.
The ideal hologram for a DoG trap being
purely real-valued, we introduce a general method based on
the Zernike phase-contrast principle to 
project real-valued holograms with the phase-only
diffractive optical elements used in standard
holographic optical trapping systems.
We demonstrate the capabilities of DoG traps
(and Zernike holograms) through experimental
studies on high-index, low-index and absorbing
colloidal particles dispersed in fluid media.
\end{abstract}

\maketitle

\section{Introduction}

Holographic optical trapping uses the forces
and torques exerted by computer-generated
holograms to manipulate microscopic objects.
Most of the literature of holographic trapping
focuses on micromanipulation of dielectric particles
with refractive indexes higher than the refractive
index of the medium, $n_p > n_m$.
Such high-index particles tend to be drawn toward
regions of high light intensity, such as the focal
point of strongly focused optical tweezers.
Low-index particles, reflecting particles and
particles that absorb light all tend to be repelled
by bright light, and therefore are difficult to
manipulate with standard optical traps.
Successful two-dimensional
manipulation of dark-seeking particles
has been achieved in the dark regions of
interference patterns
\cite{macdonald2001trapping} and the
dark core of optical vortices
\cite{gahagan1996optical,gahagan1999simultaneous}.
Full three-dimensional control has been
demonstrated
with cages of light created with rapidly
scanned optical tweezers \cite{sasaki1992optical},
and with optical bottles created from
superpositions of Bessel beams \cite{arlt2000generation,mcgloin2003three,liu2020simultaneous}.
The bright cages that define these traps all
have dark gaps through which trapped particles
can escape.
Proposals to close the gaps
have focused on the properties of
vector beams of light 
with non-trivial polarization
structure 
\cite{bokor2007three,wang2009configurable}.
These vector traps, however, cannot be projected
with standard holographic trapping systems.

Here, we report a class of optical traps
for dark-seeking particles that is based
on scalar diffraction theory and so is compatible
with standard holographic trapping techniques.
The ideal hologram encoding these traps is
entirely real-valued, which poses a challenge
for standard implementations that rely on
phase-only spatial light modulators.
We therefore introduce
a method to transform amplitude-only holograms
into phase-only holograms for convenient
projection at high diffraction efficiency.
The resulting dark tweezers can be combined with
conventional bright tweezers in a standard
holographic trapping system to enable manipulation
of heterogeneous colloidal dispersions.
We demonstrate the dark traps' capabilities
through experimental studies on
model colloidal dispersions containing
mixtures of
low-index, high-index and absorbing spheres.

\section{Dark optical tweezers}
\label{sec:darktweezers}

Conventional optical tweezers are created by bringing
a Gaussian laser beam to a diffraction-limited
focus with a high-numerical-aperture (NA) lens.
Analogous dark optical tweezers can be created
by superposing two confocal Gaussian beams with 
equal amplitudes, different waist diameters
and a relative phase of $\pi~\si{\radian}$.
The amplitude profile of such a superposition
in the focal plane
of the lens is
\begin{equation}
\label{eq:darktweezers} 
    u(\vec{r})
    =
    u_0 \left[ 
    \exp\left(-\frac{r^2}{4 \alpha^2}\right)
    -
    \exp\left(-\frac{r^2}{4 \beta^2}\right)
    \right],
\end{equation}
where $\beta$ is the radius of the dark core
and $\alpha$ is the radius of the enclosing 
region of light.
As in the case of conventional optical tweezers,
$\beta$ is constrained by the Abbe diffraction
limit to $\beta \ge \lambda/2$ for light
of wavelength $\lambda$ in the medium.
The dark trap furthermore requires
$\alpha > \beta$, with the difference
ideally exceeding $\lambda/2$.
Similar modes have been described previously
\cite{yelin2004generating} but do not appear
to have been used to create optical traps.

\begin{figure*}
    \centering
    \includegraphics[width=0.9\textwidth]{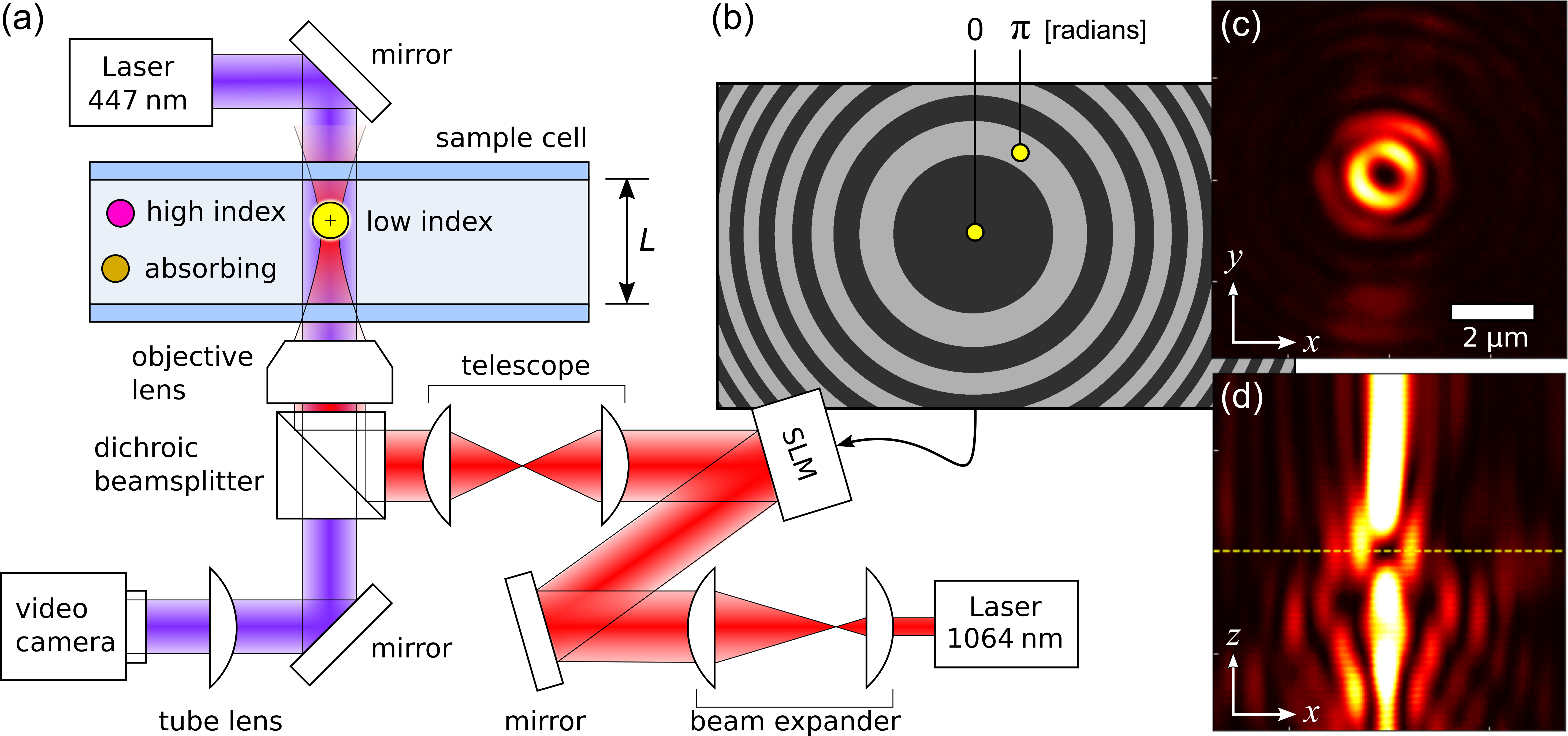}
    \caption{(a) Schematic representation of an integrated
    instrument for holographic optical trapping at \SI{1064}{\nm} 
    and holographic video microscopy at \SI{447}{\nm}. 
    The sample consists of a heterogeneous dispersion 
    of low-index colloidal spheres, 
    high-index spheres and particles that strongly
    absorb the trapping light. (b) Zernike phase hologram, $\phi_Z(\vec{r})$, 
    encoding a difference-of-Gaussians dark optical trap. 
    (c)  Measured in-plane intensity, $I(\vec{r})$, 
    at the axial position 
    indicated by the dashed line in (d).
    The extent of the dark trap in the $(x, y)$ plane 
    can be tuned by selecting the widths of the
    component Gaussian fields.
    (d) Reconstructed axial intensity profile, $I(\vec{r})$,
    projected by the hologram in (b).
    This slice in the $xz$ plane features a dark central focal volume
    completely surrounded by light.}
    \label{fig:structure}
\end{figure*}

Although the superposition described by
Eq.~\eqref{eq:darktweezers} could be implemented
with conventional optical elements, it is more
conveniently projected with a holographic optical
trapping system \cite{dufresne1998optical,grier2003revolution}
such as the example shown schematically in Fig.~\ref{fig:structure}(a).
The standard implementation imprints a 
hologram on the wavefronts of a conventional laser
using a phase-only spatial light modulator (SLM)
and then relays the modified beam to an objective
lens that focuses it into a sample.
The ideal hologram for the difference-of-Gaussians 
trap described by Eq.~\eqref{eq:darktweezers}
therefore may be computed as the Fourier
transform of the field in the focal
plane \cite{goodman2005introduction},
\begin{subequations}
\begin{align}
    h(\vec{r})
    & =
    \frac{1}{u_0}
    \int u(\vec{x}) \, 
    \exp\left(-i \frac{k}{f} \vec{r} \cdot \vec{x}\right) d^2x \\
    & = 
    \alpha^2 \exp\bigg(-\frac{k^2}{f^2}\alpha^2 r^2\bigg)
    - \beta^2 \exp\bigg(-\frac{k^2}{f^2}\beta^2 r^2\bigg),
    \label{eq:doghologram}
\end{align}
\end{subequations}
where $f$ is the focal length of the lens and 
$k = 2\pi n_m/\lambda$ is the wave number of light in
a medium of refractive index $n_m$.
Unfortunately, the hologram
in Eq.~\eqref{eq:doghologram} 
is purely real-valued and so cannot be projected
with a standard phase-only SLM.
We therefore introduce an approach inspired by
the Zernike phase-contrast technique to project
real-valued holograms with phase-only
diffractive optical elements.

\section{Zernike holograms}
\label{sec:zernikeholograms}

A complex-valued hologram 
may be factored into 
real-valued amplitude and phase profiles,
\begin{equation}
    h(\vec{r})
    =
    u(\vec{r}) \, \exp(i \phi(\vec{r})),
\end{equation}
both of which can be imprinted
onto the wavefronts of a laser beam
using suitable projection techniques
\cite{jesacher2008full,jesacher2008near-perfect}.
Most holographic trapping systems, however,
rely on phase-only diffractive optical elements
that only modify the phase profile.
A common expedient
is to ignore the
amplitude profile by setting $u(\vec{r}) = 1$, and
to imprint only the phase profile, $\phi(\vec{r})$,
onto the laser beam's wavefronts
\cite{liesener2000multi,curtis2002dynamic,curtis2005symmetry}.
The resulting phase-only hologram,
\begin{equation}
    \label{eq:truncation}
    H(\vec{r}) = \exp\left(i \phi(\vec{r}) \right),
\end{equation}
is a superposition of the ideal hologram,
$h(\vec{r})$,
with an error field,
\begin{equation}
    \Delta h(\vec{r}) 
    = 
    \left[ 1 - u(\vec{r}) \right] \,
    \exp(i \phi(\vec{r})).
\end{equation}
The effect of $\Delta h(\vec{r})$ on the
projected optical trapping pattern
depends on the complexity and symmetries
of the ideal pattern
\cite{curtis2005symmetry}.

More complicated trapping patterns tend to be
more amenable to simple phase-only
projection.
A hologram encoding multiple optical traps
can be composed by superposing  single-trap
holograms,
\begin{equation}
\label{eq:superposition}
    h(\vec{r}) = \sum_{n = 1}^N \alpha_n h_n(\vec{r}),
\end{equation}
where $h_n(\vec{r})$ is the ideal hologram for
the $n$-th trap and $\alpha_n$ is a complex coefficient
setting the relative amplitude and phase of that
trap.
An individual trap within such a pattern
can be translated by $\Delta \vec{r}$
in three dimensions
by adding a suitable parabolic profile
to the phase of its hologram
\cite{curtis2002dynamic,polin2005optimized},
\begin{equation}
\label{eq:translation}
\phi_{\Delta \vec{r}}(\vec{r})
= 
\left(
- \frac{k}{f} \vec{r}
+ \frac{k r^2}{2 f^2} \hat{z}
\right) \cdot \Delta\vec{r}.
\end{equation}
As the complexity of the trapping pattern increases,
the overall amplitude profile, $u(\vec{r})$, develops
increasingly rapid spatial variations.
The error term, $\Delta h(\vec{r})$, 
therefore tends to redirect light
away from the intended trapping pattern and
outward toward the edges of the instrument's
field of view.
This means that the phase-only hologram, $H(\vec{r})$,
can project a near-ideal rendition of the intended
trapping pattern within a limited volume.
In such cases, the
error term principally reduces the 
hologram's diffraction efficiency into the desired
mode by redirecting light elsewhere.

While the simple phase-only conversion described by
Eq.~\eqref{eq:truncation} is fast and effective, 
more sophisticated algorithms 
\cite{polin2005optimized} 
can refine a phase-only hologram to improve
diffraction efficiency and to mitigate artifacts.
These refinements typically are too slow
for real-time operation, however, and are reserved for
the most exacting applications.

Purely real-valued holograms pose a
particular challenge for phase-only projection.
Sign changes in $h(\vec{r})$
can be absorbed into the phase profile
with Euler's theorem, thereby ensuring
that the amplitude, $u(\vec{r})$,
is non-negative \cite{roichman2006projecting}.
The remaining
amplitude variations can be approximated
by multiplexing the desired hologram
with another grating to deflect light
\cite{roichman2006projecting}.
These methods either discard most of 
the light in the field of view or discard
most information about the amplitude profile.
They typically create holograms 
with undesirably low diffraction efficiency.

We improve both the fidelity and
the diffraction efficiency of both real- and 
complex-valued
holograms by transferring
more information
about the amplitude profile into the phase
profile using a variant of 
Zernike's phase-contrast
approximation,
\begin{equation}
\label{eq:zernike}  
    u(\vec{r}) 
    \approx 
    i \left[1 - \exp(i u(\vec{r}))\right],
\end{equation} 
under the assumption that $u(\vec{r}) < 1$.
Using the phase profile 
to encode amplitude information
complements Zernike's original purpose,
which was to explain how phase variations
contribute to observable contrast
in microscope images.
The phase profile implied by
Eq.~\eqref{eq:zernike},
\begin{subequations}
\label{eq:zernikehologram}
\begin{equation}
  \phi_Z(\vec{r})
  = \frac{1}{2} \, u(\vec{r}),
\end{equation}
serves as a phase-only approximation to
the real-valued amplitude profile,
yielding the phase-only Zernike approximation
to the ideal complex-valued hologram,
\begin{equation}
    H_Z(\vec{r}) = \exp\left(i \left[\phi_Z(\vec{r}) + \phi(\vec{r}) \right] \right).
\end{equation}
The associated error field,
\begin{equation}
    \label{eq:errorfieldu}
    \Delta h_Z(\vec{r}) \approx
    \left[1 - \left(1 - \frac{i}{2}\right) u(\vec{r}) \right] e^{i \phi(\vec{r})},
\end{equation}
improves upon the standard result because its
complex amplitude tends
to cancel ghost traps and other
projection artifacts.
\end{subequations}

Figure~\ref{fig:structure}(b) shows
the phase hologram encoding a dark trap
that is obtained by
applying Eq.~\eqref{eq:zernikehologram} to
the purely real-valued hologram
for a DoG trap, Eq.~\eqref{eq:doghologram}.
Although it superficially resembles
a standard Fresnel lens, the pattern
of concentric phase rings has very different
behavior.

Figure~\ref{fig:structure}(c) and \ref{fig:structure}(d) show volumetric
reconstructions \cite{roichman2006volumetric}
of the dark trap projected by the hologram
in Fig.~\ref{fig:structure}(b).
The trap is created by imprinting
the hologram on the wavefronts of
a TEM$_{00}$ laser beam at a vacuum
wavelength of \SI{1064}{\nm} (fiber laser, IPG Photonics, YLR-LP-SF) using a liquid crystal
spatial light modulator (Holoeye PLUTO).
The modified beam is relayed to the input
pupil of an objective lens
(Nikon Plan Apo, $100\times$,
numerical aperture 1.4, oil immersion)
that focuses the light into the intended
optical trap with a focal length
of $f = \SI{180}{\um}$.
The trapping beam is diverted into the
objective lens with a dichroic beamsplitter
(Semrock) that has a reflectivity
of \SI{99.5}{\percent} at the trapping
wavelength.

\begin{figure*}
    \centering
    \includegraphics[width=0.75\textwidth]{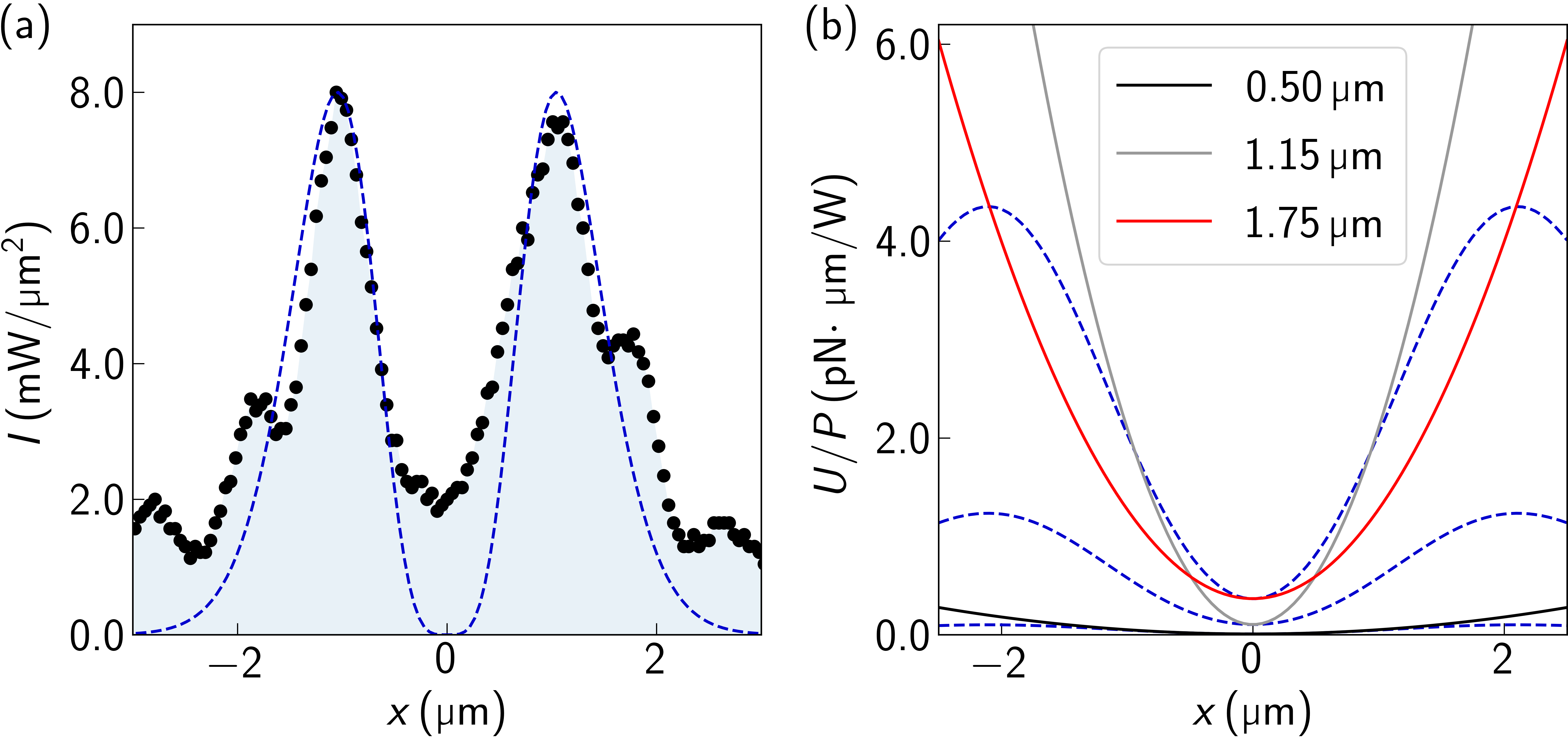}
    \caption{(a) Measured intensity (discrete points) 
    of a DoG trap with
    $\alpha = \num{0.96}$ and $\beta = \alpha/2$
    measured in the plane indicated by the horizontal
    line in Fig.~\ref{fig:structure}(d). 
    The dashed blue curve 
    is the ideal intensity distribution, $\abs{u_0(x)}^2$,
    for this trap.
    (b) Dashed curves show the potential energy computed
    with Eq.~\eqref{eq:potentialmodel} for three different
    sizes of silica spheres dispersed in DMSO solution
    and localized in the trap from (a). Solid curves
    show the associated Hookean wells computed 
    from the experimentally measured stiffness of the
    trap for each of the particles.}
    \label{fig:potential}
\end{figure*}

Images of the projected intensity pattern are
obtained by mounting a front-surface
mirror in the focal plane of the objective lens \cite{roichman2006volumetric}.
The reflected light is collected by the
objective lens, and a small proportion
passes through the dichroic mirror.
This transmitted light is collected
with a \SI{200}{\mm} tube lens and is
recorded with a video camera (FliR Flea3,
monochrome) with an effective system
magnification of \SI{0.048}{\um\per pixel}.
Transverse intensity slices, such as
the example in Fig.~\ref{fig:structure}(c),
are obtained by translating the trap
in steps of $\Delta z = \SI{48}{\nm}$
along the axial direction using
Eq.~\eqref{eq:translation}.
A stack of slices then is combined
to obtain the axial section that is
presented in Fig.~\ref{fig:structure}(d).
These images
confirm that the Zernike hologram for
a DoG trap successfully projects a 
beam of light that focuses to a dark
volume surrounded by light on all sides.
Planned and measured intensity profiles
are compared in Fig.~\ref{fig:potential}(a)
for a trap with $\alpha = \SI{0.96}{\um}$
and $\beta = \alpha/2$.
This trap is used for experimental validation
measurements in Section~\ref{sec:performance}.

\section{Theoretical performance}
\label{sec:theory}

\begin{widetext}
\begin{subequations}
\label{eq:potentialmodel}
A dark-seeking particle at distance $r$
from the focus of
a DoG trap has a potential energy
that depends on its overlap with the light.
We model this overlap as
\begin{align}
    U(r) 
    & =
    A \, 
    \int_0^{2 \pi} d\theta \,
    \int_0^{a_p} dx \,
    x \, \zeta(x) \, u^2(\vec{x} - \vec{r}) \\
    & =
      4 \pi a_p^3 u_0^2 \, A
      \left[
    g\left(r, a_p, \alpha^2\right) 
    + g\left(r, a_p, \beta^2\right)
      - 2 g\left(r, a_p, \frac{2 \alpha^2\beta^2}{\alpha^2 + \beta^2}\right)
    \right] ,
\end{align}
where $u(\vec{r})$ is the DoG trap field from Eq.~\eqref{eq:darktweezers} and
$\zeta(x) = 2 \sqrt{a_p^2 - x^2}$ 
is the chord length through a sphere of radius
$a_p$ at a distance $x$ from its center.
This solution depends on the overlap
integral
\begin{equation}
\label{eq:g}
    g\left(r, a_p, \alpha^2\right)
    = 
    e^{-\frac{r^2}{2 \alpha^2}}
    \int_0^1 y \sqrt{1 - y^2} \, 
    I_0\left(\frac{a_p r}{\alpha^2} \, y \right)\,
    e^{-\frac{a_p^2}{2 \alpha^2} \, y^2} \, dy,
\end{equation}
whose integrand depends on
the modified Bessel function, $I_0(\cdot)$.
We treat the overall scale of the trapping potential, $A$, as an adjustable parameter.
This constant accounts phenomenologically
for the contrast in refractive index
between the sphere and the medium as well
as the effect of refraction within the
sphere, details of which are beyond
the scope of this model.
Numerical solutions for the dark trap's
potential energy landscape are plotted
in Fig.~\ref{fig:potential}(b) for three different particle
sizes in a trap with $\alpha = \SI{0.96}{\um}$
and $\beta = \alpha/2$.
\end{subequations}

\begin{subequations}
\label{eq:qualitymodel}
As expected, DoG traps have a potential-energy minimum 
at $r = 0$.
The transverse trapping stiffness therefore
can be obtained from Eq.~\eqref{eq:potentialmodel} as
\begin{align}
    \kappa_\perp(a_p, \alpha, \beta)
    & \equiv
    \lim_{r \to 0} \frac{\partial^2 U}{\partial r^2} \\
    & =
    2 \pi a_p \, u_0^2 A \left[
    f\left(\frac{a_p}{\alpha}\right) 
    + f\left(\frac{a_p}{\beta}\right)
    - 2 f\left(\frac{a_p}{\alpha\beta} 
    \sqrt{\frac{\alpha^2 + \beta^2}{2}} \right)
    \right], 
\end{align}
which depends on the Dawson integral, $F(\cdot)$, through
\begin{equation}
    \label{eq:f}
    f(x) 
    = 
    1 - 
    \sqrt{2} \left(x + \frac{1}{x} \right) \, F\left(\frac{x}{\sqrt{2}}\right).
\end{equation}
The associated trapping efficiency is
\begin{equation}
    \label{eq:q}
    Q_\perp(a_p, \alpha, \beta) 
    = \frac{1}{P} \, \kappa_\perp(a_p, \alpha, \beta),
\end{equation}
where $P \propto u_0^2$ is the laser power delivered
to the trap.
For the particular choice of $\beta = \alpha/2$,
the optimal value of the trap width is
$\alpha^\ast = \num{0.952} \, a_p$.
The corresponding
optimal efficiency,
$Q(a_p, \alpha^\ast, \alpha^\ast/2) = \num{1.44} \, a_p A$,
depends on the relative refractive index
of the particle through the phenomenological
scale factor, $A$.
Predictions from Eq.~\eqref{eq:qualitymodel}
can be compared with the measured performance
of DoG traps for low-index dielectric particles.
\end{subequations}
\end{widetext}

\section{Measured performance}
\label{sec:performance}

The performance of an optical trap can be assessed
by tracking thermally-driven fluctuations in 
a trapped particle's position \cite{florin1998photonic}.
We measure these fluctuations using the
instrument's holographic microscopy subsystem,
as illustrated
schematically in Fig.~\ref{fig:structure}.
In-line holograms of the particles
are recorded by
illuminating the sample with the
collimated beam from a diode laser
(Coherent Cube) operating at a vacuum
wavelength of $\lambda = \SI{447}{\nm}$.
Light scattered by a particle 
interferes with the
rest of the beam in the focal plane of
the microscope.
The intensity of the magnified
interference pattern is recorded by the
video camera at
\SI{30}{frames\per\second}.
The camera's \SI{10}{\us} exposure time is short enough to
avoid motion blurring \cite{cheong2009flow,dixon2011holographic}.
A holographic snapshot can be analyzed 
with predictions of Lorenz-Mie theory
to measure each particle's radius, $a_p$,
refractive index, $n_p$ and
three-dimensional position, 
$\vec{r}_p = (x_p, y_p, z_p)$, relative
to the center of the microscope's
focal plane \cite{lee2007characterizing}.
The typical precision for such holographic
characterization and tracking measurements is
$\sigma_{a_p} = \SI{+-2}{\nm}$,
$\sigma_{n_p} = \num{+-1e-3}$,
$\sigma_{x_p} = \sigma_{y_p} = \SI{+-2}{\nm}$
and $\sigma_{z_p} = \SI{+-5}{\nm}$
\cite{krishnatreya2014measuring}.
We use these capabilities to differentiate
high-index, low-index and absorbing particles
on the basis of their refractive indexes
and to track their three-dimensional motions
as they are manipulated in holographic
optical traps.

\begin{subequations}
\label{eq:stiffnessestimator}
Treating a sphere's thermal fluctuations within
the trap as an Ornstein-Uhlenbeck process, an $N$-step
discretely sampled trajectory, $\{x_n\}$, along $\hat{x}$
yields estimates \cite{polin2005optimized}
for the associated stiffness,
\begin{equation}
    \label{eq:stiffnessestimate}
    \kappa_x = \frac{k_B T}{c_0},
\end{equation}
and its uncertainty,
\begin{equation}
    \label{eq:stiffnessuncertainty}
    \Delta \kappa_x
    =
    \frac{k_B T}{c_0}
    \sqrt{\frac{2}{N} \left(1 + \frac{2 c_1^2}{c_0^2 - c_1^2}\right)} ,
\end{equation}
where $k_B T$ is the thermal energy scale at absolute
temperature $T$ and
\begin{equation}
    c_j = \frac{1}{N} \sum_{n = 1}^N x_n x_{(n+j) \bmod N}
    \label{eq:correlators}
\end{equation}
is the autocorrelation of $\{x_n\}$ at lag $j$.
Similar estimates can be computed for the trap stiffness
along the other Cartesian coordinates.
Equation~\eqref{eq:stiffnessestimator} is suitable
for interpreting holographically measured single-particle
trajectories because the typical scale
of thermal fluctuations, 
$\Delta x = \sqrt{2(c_0 - c_1)} \ge \SI{70}{\nm}$,
is much greater than the measurement error,
$\sigma_x \le \SI{5}{\nm}$,
over the range of laser powers considered
\cite{polin2005optimized}.
A \num{2000}-frame holographic video 
recorded over one minute
provides enough information 
to measure the trap stiffness, $\kappa_x$, to within 
\SI{1}{\percent} in each of the three dimensions.
\end{subequations}

\begin{figure*}
    \centering
    \includegraphics[width=0.9\textwidth]{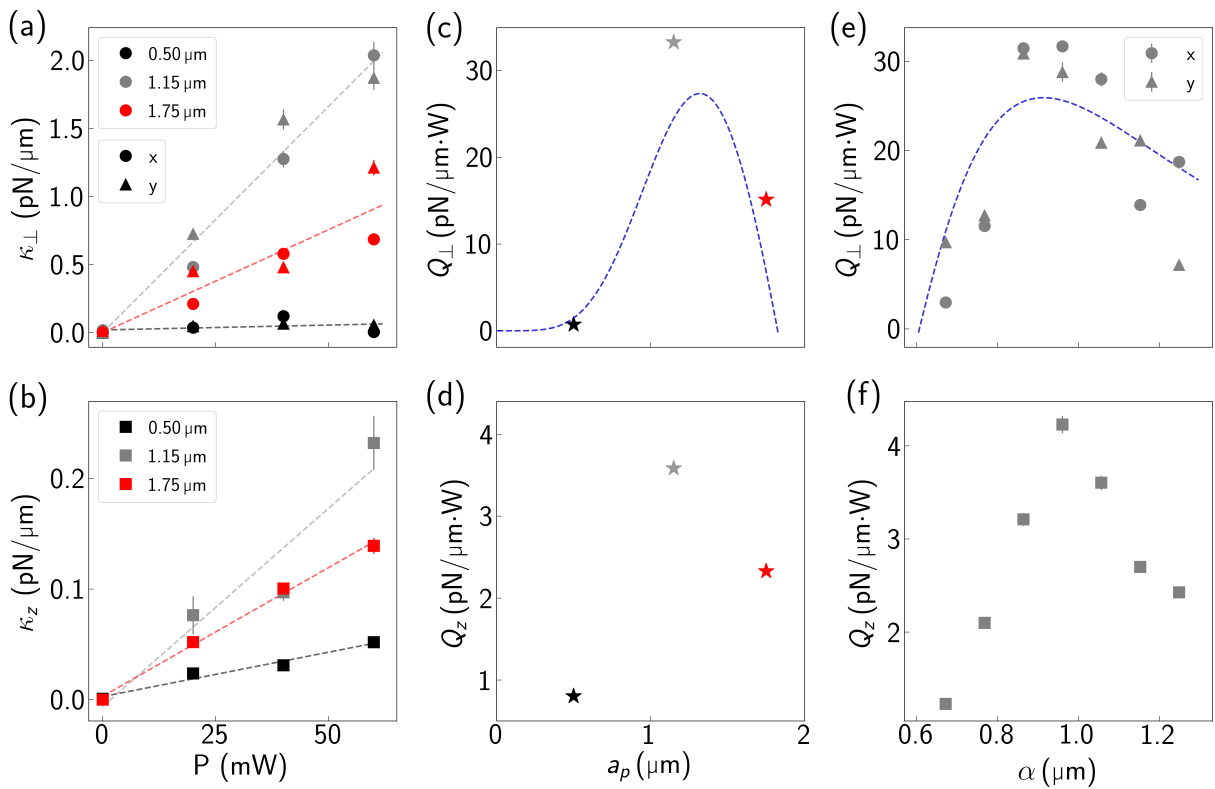}
    \caption{
    (a) Transverse stiffness
    $\kappa_\perp(\alpha, a_p)$,
    of a DoG trap with $\alpha = \SI{0.96}{\um}$ and
    $\beta = \alpha/2$ as a function of laser power.
    Results obtained from thermal fluctuations in
    the $(x,y)$ plane are plotted for silica spheres of
    radii $a_p = \SI{0.5}{\um}$, \SI{1.15}{\um} and
    \SI{1.75}{\um}.
    (b) Power dependence of the axial stiffness, $\kappa_z(\alpha, a_p)$.
    (c) Size dependence of the transverse trapping efficiency, $Q_\perp(\alpha, a_p)$, for fixed
    $\alpha$.
    (d) Size dependence of the axial trapping
    efficiency, $Q_z(\alpha, a_p)$.
    (e) Transverse trapping efficiency as a function
    of trap shape parameter, $\alpha$, 
    for a silica sphere with
    $a_p = \SI{1.15}{\um}$.
    The blue curves in (c) and (e) are a comparison with
    predictions of Eq.~\eqref{eq:qualitymodel}.
    (f) Axial trapping efficiency as a function of
    $\alpha$.}
    \label{fig:stiffness}
\end{figure*}

Figure~\ref{fig:stiffness}(a) shows how the 
measured stiffness of the DoG trap from Fig.~\ref{fig:potential}(a)
depends on laser power
and particle radius for colloidal
silica spheres dispersed in an aqueous solution of
dimethylsufoxide 
(DMSO, Acros Organics, CAS Number 67-68-5). 
Silica has a refractive index of
$n_p = \num{1.42}$.
The DMSO solution has 
a refractive index of
$n_m = \num{1.53}$, as measured by an Abbe
refractometer (Edmund Scientific, Model 52-975).
The silica spheres therefore have a lower
refractive index than their medium and cannot
be trapped with conventional optical tweezers.
The data in Fig.~\ref{fig:stiffness}(a) and (b) 
were acquired
for three particle sizes,
$a_p = \SI{0.505}{\um}$ (Polysciences Inc., 24326),
\SI{1.15}{\um} (Bangs Laboratories, SS04001) and
\SI{1.75}{\um} (Bangs Laboratories, SS05001).
The diameters and refractive indexes of these
particles were confirmed \emph{in situ} through
holographic particle characterization
\cite{lee2007characterizing}.

The DoG trap used for these measurements
has $\alpha = \SI{0.96}{\um}$ and 
$\beta = \alpha/2 = \SI{0.48}{\um}$,
which most closely matches the optimal width
for the particles with radius $a_p = \SI{1.15}{\um}$.
The power, $P$, projected into the trap is measured with 
a slide-mounted thermal power sensor (Thorlabs model S175C).
As might reasonably be expected, the trap stiffness increases 
linearly with $P$ in all three
coordinates for each of the three particle sizes.
Values for the transverse stiffness plotted in Fig.~\ref{fig:stiffness}(a) 
reveal no anisotropy, and thus no dependence on the trap's polarization,
which is directed along $\hat{x}$.
This is consistent with the performance of conventional optical tweezers for light-seeking particles
that are larger than the wavelength of light \cite{rohrbach2005stiffness,madadi2012polarization}.
Also like conventional optical traps, the axial stiffness of DoG traps, plotted in Fig.~\ref{fig:stiffness}(b), 
differs significantly from the transverse stiffness.

Figures~\ref{fig:stiffness}(c) and \ref{fig:stiffness}(d) show
how the transverse and axial trapping efficiencies, $Q_\perp(a_p, \alpha) = 
[\kappa_x(a_p, \alpha) + \kappa_y(a_p, \alpha)]/(2P)$
and $Q_z(a_p, \alpha) = \kappa_z(a_p, \alpha)/P$, 
depend on particle radius, $a_p$.
Interestingly, the measured transverse trapping
efficiency, $Q_\perp(a_p, \alpha)$, 
depends non-monotonically
on particle size.
This observation is consistent with the prediction
of Eq.~\eqref{eq:qualitymodel}, which is plotted
as a blue dashed curve in Fig.~\ref{fig:stiffness}(c),
using only $A$ as an adjustable parameter.
The model predicts that the DoG trap
is optimally stiff when trapping 
low-index dielectric particles that are
slightly larger than trap's dark core,
in agreement with experimental observations.

The complementary results in Figs.~\ref{fig:stiffness}(e) and \ref{fig:stiffness}(f) show how the trapping
efficiency depends on $\alpha$ for fixed particle
size, $a_p = \SI{1.15}{\um}$.
This result illustrates in more detail how the trapping
efficiency of DoG traps is optimized when the trap
is designed to fit the particle.
This kind of optimization is facilitated by the
combining holographic trapping with
real-time holographic particle characterization.
The peak trapping efficiency
around \SI{30}{\pico\newton\per\um\per\watt} for a particle with
size parameter
$k a_p = 10$ and relative
refractive index $n_p/n_m =
\num{0.93}$ is comparable
to the computed efficiency
of a conventional optical tweezer
for a complementary high-index particle with relative
refractive index 
$n_p/n_m = (0.93)^{-1} = \num{1.07}$
\cite{nieminen2007optical}.

\section{Simultaneous manipulation of dark- and light-seeking particles}

\begin{figure}[b]
    \centering
    \includegraphics[width=\columnwidth]{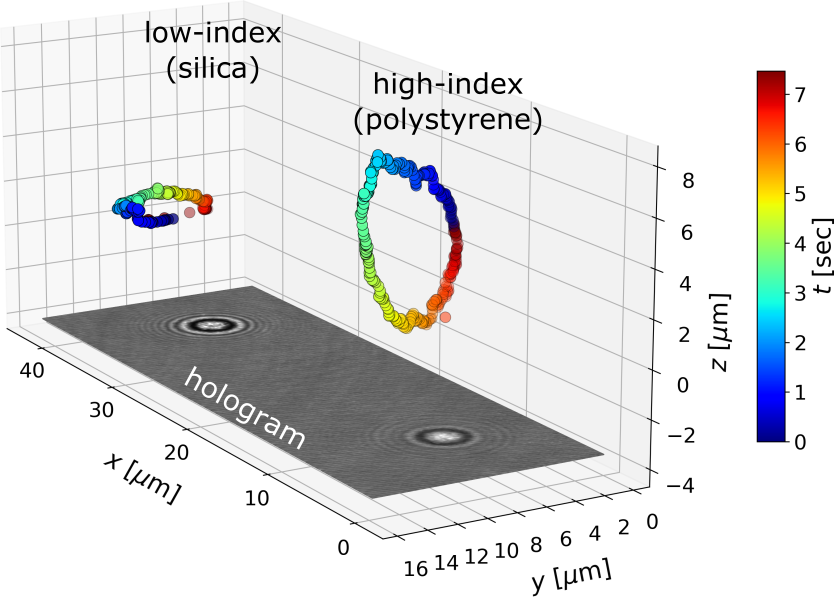}
    \caption{A low-index (silica) particle and 
    a high-index (polystyrene) particle are trapped simultaneously in
    a DoG trap and an optical tweezer, respectively, and are transported around 
    nominally circular paths in the vertical $yz$ plane.
    Discrete points show the particles' actual positions as measured
    by holographic particle tracking applied to Supplementary
    Video 1, and are colored by time.
    The gray-scale image is one
    hologram from that video.}
    \label{fig:trajectories}
\end{figure}

Figure~\ref{fig:trajectories} 
presents holographically measured trajectories
of high- and low-index particles being translated
simultaneously in a conventional optical tweezer
and a DoG trap, respectively.
The figure also includes one frame from the holographic
video (Supplementary Video 1) that was used to measure
the particles' trajectories.
The two types of particles are co-dispersed in
an aqueous DMSO solution with a measured
refractive index of $n_m = \num{1.53}$.
The low-index particle is a silica sphere
with a radius of $a_p = \SI{1.15}{\um}$,
and a holographically measured refractive index of
$n_p = \num{1.42}$.
The high-index particle is composed
of polystyrene with a radius of $a_p = \SI{0.5}{\um}$ 
(ThermoFisher Scientific, 5100B) and
a refractive index, $n_p = \num{1.60}$, that
exceeds the index of the medium.
The dispersion
is contained in the volume created
by sealing the edges of
a glass coverslip to the
face of a glass microscope slide.
This chamber has an optical path length
of $L = \SI{25}{\um}$.

The discrete plot symbols in Fig.~\ref{fig:trajectories}
represent the measured three-dimensional
trajectories of a (low-index) silica sphere
and a (high-index) polystyrene
sphere as they are transported in opposite
directions around nominally circular
trajectories in the vertical $yz$ plane.
The particles follow the programmed
trap trajectories at
\SI{5.0}{\um\per\second} 
given a laser power of
\SI{100}{\milli\watt} for
the conventional optical trap and
\SI{50}{\milli\watt} for
the DoG trap as estimated by imaging
photometry \cite{roichman2006volumetric}.
Comparatively high power is needed to manipulate
the polystyrene particle because it is nearly
index matched to the medium.
The traps' relative intensities are adjusted by appropriately weighting each
trap's contribution to the superposition in
Eq.~\eqref{eq:superposition}.

Although the two traps were programmed to move
in circles of equal radius, the recorded trajectories
are elliptical.
The two particles move over equal
horizontal ranges, but the
polystyrene particle moves further than
expected in the axial direction and the
silica particle moves less far.
Similar axial deviations have been reported
for particles in conventional
optical tweezers \cite{obrien2019above}
and can be ascribed to differences in the
particles' buoyant densities and to changes
in the traps' Rayleigh ranges as they
are displaced along the optical axis.
These deviations can be corrected through
adjustments to the axial displacements in
Eq.~\eqref{eq:translation}
using feedback from real-time holographic tracking.

\section{Manipulating absorbing particles}
\label{sec:absorbing}

DoG traps also are useful for manipulating
particles that absorb light strongly.
Such particles are repelled by conventional
optical tweezers through transfer
of momentum from absorbed light.
Absorption also mediates heating, which
can propel the particles through
self-thermophoresis \cite{shvedov2010giant,moyses2016trochoidal},
and in extreme cases can destroy the
particles or boil the fluid medium
\cite{rubinszteindunlop1998optical,peterman2003laser}.

We have demonstrated the ability of DoG
traps to stably trap and transport absorbing
particles through experiments on
composite particles made of hematite
cubes embedded in dielectric
spheres and
superparamagetic particles composed of
hematite nanoparticles dispersed within
dielectric spheres.
The dielectric spheres used for these demonstrations
are created by emulsion polymerization of 
3-methacryloxypropyl trimethoxysilane (TPM)
\cite{vanderwel2017preparation},
an organosilicate with a refractive index
of $n_p = \num{1.495}$
\cite{middleton2019optimizing}.
Hematite absorbs infrared light strongly,
and both types of hematite-loaded particles
tend to be ejected from conventional 
optical tweezers.
Both types of particles are readily
trapped and transported in three dimensions
with DoG traps.
Supplementary Video 2 shows the simultaneous
three-dimensional control over a 
$\SI{1}{\um}$-diameter polystyrene particle
and a 
\SI{1.8}{\um} TPM particle enclosing a
\SI{0.8}{\um} hematite cube.

\section{Discussion}
\label{sec:discussion}
This study introduces a class of ``dark" optical
traps that are created
by superposing two Gaussian modes of different 
waist diameters and opposite phases.
The hologram encoding these Difference-of-Gaussians
(DoG) traps is purely real-valued.
We therefore introduce a technique based on
the Zernike phase-contrast approximation
to project real-valued holograms with
phase-only diffractive optical elements,
such as the spatial light modulators
commonly used in holographic optical trapping systems.
We have demonstrated DoG traps' ability
to trap dark-seeking particles and to move them
in three dimensions.
We further have characterized the efficiency
of DoG traps for localizing
low-index particles and have explained the
observed nonmonotonic dependence of the
transverse trapping efficiency on particle size.
When a DoG trap is optimally matched to the
trapped particle, its trapping efficiency is
comparable to that of conventional optical
tweezers for light-seeking particles.

\section*{Funding}
This work was supported by the National Science Foundation
through Award No.\ DMR-2104836.
The integrated instrument for holographic
trapping and holographic microscopy used
in this study was constructed with support of the MRI program of the NSF under Award Number DMR-0922680 and
is maintained as shared instrumentation by the
Center for Soft Matter Research at NYU.
The authors gratefully acknowledge the support of the 
nVidia Corporation through the donation of the
Titan Xp and Titan RTX GPUs used for this research.

\section*{Acknowledgments.}
We are grateful to Shahrzad Zare, who contributed
to the initial stages of this work.

\section*{Data Availability.}
Data underlying the results presented in this paper
are not publicly available at this time but may be
obtained from the authors upon reasonable request.

The open-source software used to project
holographic optical traps, record
in-line holographic microscopy data
and analyze those data is available
online at
\url{https://github.com/davidgrier/}.


\end{document}